\documentclass[11pt,epsfig,twoside]{article}
\usepackage{xspace}
\usepackage{relsize}

\def\babar{\mbox{\slshape B\kern-0.1em{\smaller A}\kern-0.1em
    B\kern-0.1em{\smaller A\kern-0.2em R}}}

\def\cp{$CP$}
\def\rs{\raisebox{1.5ex}[-1.5ex]}
\def\ifb{fb^{-1}}
\def\B{$B$}
\def\de{\Delta E}
\def\mes{\ensuremath{m_{\rm ES}}}
\def\btopp{$B \to \pi\pi$}

\def\btohhh{$B \to h h h$}
\def\btorr{$B \to \rho \rho$}
\def\btokstr{$B \to K^* \rho$}
\def\btosgl{$b\to s ~{\rm gluon}$}
\def\btosdg{$b\to s \gamma/d \gamma$}
\def\btosg{$b\to s \gamma$}
\def\btoeprkst{$B\to \eta' K^*$}
\def\btopkst{$B\to\phi K^*$}
\def\btokstg{$B\to K^* \gamma$}
\def\btorg{$B\to \rho \gamma$}
\def\btoszgll{$b \to s Z(l\bar{l})/s \gamma (l \bar{l})$}
\def\btoknn{$B\to K\nu\bar{\nu}$}
\def\ups{\Upsilon(4S)}
\def\MeVcc{${\rm  MeV}/c^2$}
\def\GeVcc{${\rm  GeV}/c^2$}

\def\GeVc{{\rm  GeV/c}}
\def\MeVc{{\rm  MeV/c}}

\def\MeVcd{{\rm  MeV/c^2}}
\def\Bz{\ensuremath{B^0}\xspace}
\def\Bzb  {\ensuremath{\Bbar^0}\xspace}
\def\BBbar{\ensuremath{\Bz {\kern -0.16em \Bzb}}\xspace}
\def\BB{\ensuremath{B {\kern -0.16em \overline{B}}}\xspace}
\def\Bbar {\kern 0.18em\overline{\kern -0.18em B}{}\xspace}
\def\cosBthr{{\rm cos}\theta(T_B,z)}
\def\roe{{\rm ROE}}

\def\Journal#1#2#3#4{{#1} {\bf #2}, #3 (#4)}

\def\NIMA{{\em Nucl. Instrum. Methods} A}

\def\PLB{{\em Phys. Lett.}  B}

\def\PRD{{\em Phys. Rev.} D}

\def\ie{{\em i.e.}} 
\def\cf{{\em cf.}} 
 
\def\ea{{\em et al.}}

\def\mkp{$m_{K\pi}$}
\def\mpp{$m_{\pi\pi}$}
\begin{document}
\begin{titlepage}
\setcounter{page}{1}

\begin{flushright} 
LAL 03-53 \\
\end{flushright} 

\begin{center}
\vspace{2.5cm}
{\large\bf
RARE B DECAYS AND DIRECT CP VIOLATION AT \babar
}
\vspace{1.5cm}
\begin{large}
Sandrine~Laplace \\
\end{large}
\vspace{0.5cm}
\begin{large}
On behalf of the \babar\ Collaboration\\
\end{large}
\vspace{1.cm}
{\small \em Laboratoire de l'Acc\'el\'erateur Lin\'eaire,\\
IN2P3-CNRS et Universit\'e de Paris-Sud, BP 34, F-91898 Orsay Cedex, France}\\
\vspace{3.5cm}

{\small{\bf Abstract}}
\end{center}
{
\vspace{-0.2cm}
The search for rare \B\ decays and direct \cp\ violation at \babar\ is described. 
The following measurements (based on integrated luminosities ranging from $56.4$ to $81.9~\ifb$)
are summarized: the inclusive branching fractions and direct \cp\ asymmetries
of $B^+\to h^+h^-h^+$ ($h=\pi, K$), the exclusive branching fractions 
of $B^+\to K^+\pi^-\pi^+$ (where significant signals are observed in the
$B^+ \to K^{*0}(892)\pi^+$, $B^+ \to f_0(980)K^+$, $B^+ \to \chi_{c0}K^+$,
$B^+ \to \overline{D}^0\pi^+$ and $B^+ \to {\rm higher}~K^{*0}\pi^+$ channels),
the branching fractions of $B^+\to \rho^0\rho^+$ and $B^+\to \rho^0 K^{*+}$,
and finally, the branching fractions, the longitudinal components, and the
direct \cp\ asymmetries in $B\to \phi K^*$.
}

\vspace{3.5cm}
\centerline{\small\em To appear in the proceedings of the 17th Les Rencontres de Physique de la Vallee d'Aoste: }
\centerline{\small\em Results and Perspectives in Particle Physics, La Thuile, Aosta Valley, Italy, 9-15 Mar 2003}
\vspace{1cm}
\thispagestyle{empty}

\end{titlepage}

\newpage

\section{Introduction}

Measurements of the branching fractions and direct \cp\ asymmetries of rare \B\ decays using the \babar\ 
detector\cite{babardet} are presented. 
Rare \B\ decays can be classified according to their suppression factor:
\begin{itemize}
\item {\bf a small CKM matrix element}, as for charmless \B\ decays which amplitudes
are suppressed by a factor $|V_{ub}|/|V_{cb}|\simeq \lambda$ compared to charm \B\ decay 
amplitudes. Measurements related to the decays \btohhh ($h=\pi,~K$), \btorr\ and \btokstr\ are
presented. 
\item {\bf a dominant diagram involving a loop}, which amplitudes are suppressed by either 
\begin{itemize}
\item $\alpha_S/4\pi$ for gluonic penguin (\btosgl) as in the exclusive decays \btoeprkst\ and \btopkst.
\item $\alpha_{\rm QED}/4\pi$ for radiative penguins (\btosdg), as in the exclusive decays 
\btokstg\ and \btorg\ or the inclusive decay \btosg, and for electroweak penguins (\btoszgll),
as in the exclusive decay \btoknn. 
\end{itemize}
\end{itemize}

All the measurements mentioned above were presented in my talk, but 
only the results newly released during the winter conferences
are detailed in these proceedings, \ie, \btohhh ($h=\pi,~K$), \btorr, \btokstr\ and \btopkst. 

The motivation for such measurements is twofold: first, they could allow to see indirect
effects of new physics particles virtually created in loop. Such effects may show up in different
branching fractions and \cp\ asymmetries than those predicted by the Standard Model.
One concentrates here on time-integrated direct \cp\ asymmetries. Time-dependent asymmetries
are described elsewhere\cite{faccini}.

Time-integrated direct \cp\ asymmetries require at least two (Standard Model or New Physics)
amplitudes contributing to a process with different weak and strong phases: calling
$A$ the total amplitude of the process $B^0~(B^+)\to f$ and $\bar{A}$ the one
of the \cp\ conjugated process $\overline{B}^0~(B^-)\to f$, one can split the amplitudes
$A$ and $\bar{A}$ into a sum of real amplitudes $A_k$, \cp-odd weak phases $\phi_k$
and \cp-even strong phases $\delta_k$:
\begin{equation}
A = \sum_k A_k e^{i\phi_k} e^{i\delta_k}, \quad \bar{A} = \sum_k A_k e^{-i\phi_k} e^{i\delta_k}.
\end{equation}
The asymmetry between $A$ and $\bar{A}$ can then be written in terms of differences of weak and strong
phases (the expression below is given in the case where two amplitudes contribute to the process):
\begin{equation}
|A|^2 - |\bar{A}|^2 = - 4 A_1 A_2 \sin(\phi_1-\phi_2) \sin(\delta_1-\delta_2). 
\end{equation}

Beyond the search for hints of new physics, measurements of rare \B\ decays can help
constraining the Standard Model unknown parameters. For example, time-dependent \cp\ asymmetries
in \btopp\ and \btoeprkst, \btopkst\ allow to measure $\sin2\alpha$ and $\sin2\beta$
where $\alpha$ and $\beta$ are two angles of the Unitarity Triangle; the ratio
of the \btorg\ and \btokstg\ rates constrains the ratio of CKM matrix elements $|V_{td}|/|V_{ts}|$;
finally, the \btosg\ spectrum helps to measure $V_{ub}$ and to
constrain Heavy Quark Effective Theory parameters. 

\section{The \babar\ detector and dataset}

The \babar\ detector is described elsewhere\cite{babardet} in detail. It consists of a tracking
system composed from a 5-layer double sided silicon micro strip vertex tracker (SVT) and from
a 40-layer drift chamber (DCH), both operating in a $1.5$ T solenoidal magnetic field. 
Charged particle identification is mainly performed using a ring imaging Cherenkov detector (DIRC): the separation
between kaons and pions ranges between $8\sigma$ for a momentum of $2~{\rm GeV/c^2}$ and $4\sigma$
for a momentum of $4~{\rm GeV/c^2}$. Photons and neutral hadrons are detected in a CsI(Tl) electromagnetic calorimeter.
The identification of muons and neutral hadrons is done in the flux return instrumented with 
many layers of resistive plate chambers (IFR). 

The data used in the analyses presented here were collected between 1999 and 2002. The total
luminosity integrated at the $\ups$ resonance (on-resonance data) is $81.9~\ifb$, and the one
integrated $40$ MeV below the $\ups$ resonance (off-resonance data, used for continuum background studies)
is $9.6~\ifb$. Some of the analyses are performed on a fraction of the total integrated luminosity
only.

\section{Kinematics and event topology at a \B\ factory: background fighting}

At the PEP-II \B\ factory, \B\ mesons are produced by colliding electrons and positrons
at a center of mass energy equal to the $\ups$ mass.
At this energy, the $b\bar{b}$ cross-section corresponds to about $25\%$
of the total cross-section (including as well lighter quark production, called 
thereafter continuum). Continuum is a first source of background. Cross-talk
from other \B\ decays is a second source of background referred to as \B\
background in the following.

The kinematic and topological variables described in this section
are used to distinguish signal \B\ events from continuum and \B\ backgrounds.

\subsection{Kinematic variables}

The conservation of energy and momentum allows us to build the
following two (nearly uncorrelated) kinematic variables:
\begin{itemize}
\item The variable $\de$ is defined as:
\begin{equation}
\de = E^*_B - \sqrt{s}/2,
\label{eq:de}
\end{equation}
where $E^*_B$ is the \B\ candidate energy and $\sqrt{s}$ is the beam energy, both
calculated in the $\ups$ center of mass. When the \B\ candidate corresponds to a real
\B\ decay, $\de$ is close from zero, up to the resolution which is dominated by the reconstruction
of the \B\ energy (a few tens of MeV, depending on the nature of the daughters of the \B). 
\item the energy-substituted mass, $\mes$, is defined by:
\begin{equation}
\mes = \sqrt{(s/2 + \vec{p}_i.\vec{p}_B)^2/E_i^2-\vec{p}_B^2},
\label{eq:mes}
\end{equation}
where $E_i$ and $\vec{p}_i$ are respectively the total energy and the momentum of the
$e^+e^-$ pair in the laboratory frame, and $\vec{p}_B$  is the momentum of the reconstructed
\B\ candidate. For a real \B\ decay, $\mes$ peaks around the \B\ mass, up to the resolution
of $2.6~\MeVcd$ dominated by the beam energy dispersion. 
\end{itemize}

Kinematic variables are used both for continuum and \B\ background
rejection. Because they are mostly uncorrelated, these two variables are often combined
into a likelihood function.

\subsection{Topological variables}

At the $\ups$ energy, $\BBbar$\ pairs are produced almost at rest
in the $\ups$ center of mass. Therefore, in this frame, the \B\ daughters are 
isotropically distributed. Oppositely, for the lighter quarks (mainly for the $u,d,s$ quarks, 
to a lesser extend for the $c$ quark) extra energy is available to boost the produced
particles, leading to a back-to-back jet structure. 

Moreover, in the process $e^+e^-\to\ups\to \BB$ where the spin-$1$ $\ups$ decays into
two spin-$0$ \B\ mesons, the angular distribution of the \B\ in the center of mass follows
a $\sin^2\theta_B$ distribution where $\theta_B$ is the angle between the \B\ direction and the
beam axis. Contrarily, in the process $e^+e^-\to f\bar{f}$ (where $f$ is a fermion), the distribution follows
$1+\cos^2\theta_T$ where $\theta_T$ is the angle between the ``jet direction'' and the beam axis.

Topological variables can be built to take advantage of these shape properties. Amongst them,
one can retain:
\begin{itemize}
\item $\cosBthr$: the cosine of the angle between the \B\ candidate 
thrust direction and the beam axis. True \B\ decays lead to a flat distribution, while
one retrieve the $1+\cos^2\theta$ distribution mentioned earlier for continuum
background. 
\item $L_0$ and $L_2$: momentum weighted monomials defined as:
\begin{equation}
\label{eq:monomialsApp}
L_n=\sum_{i=\roe} p_i |{\rm cos}(\theta_{T_B,i})|^n~,
\end{equation}
where ${\rm cos}(\theta_{T_B,i})$ is the cosine between the thrust direction of the \B\
candidate and the $i$-th track of the Rest Of the Event, ROE, (corresponding to what
remains in the event once the \B\ candidate tracks are removed).
\end{itemize}

Because topological variables are related to \B\ decays and continuum production properties,
they are mostly used for continuum background rejection, while being inefficient for
\B\ background rejection.

Often highly correlated amongst each others, they are combined using Fisher\cite{Fisher}
or neural network discriminants. Figure~\ref{fig:fisher} shows the Fisher
discriminant outputs using the variables $L_0$ and $L_2$ for
$B^0\to\pi^+\pi^-$ Monte Carlo and continuum background. One obtains a good 
separation between these two categories of events. 

\begin{figure}[t]
\vspace{8.5cm}
\includegraphics{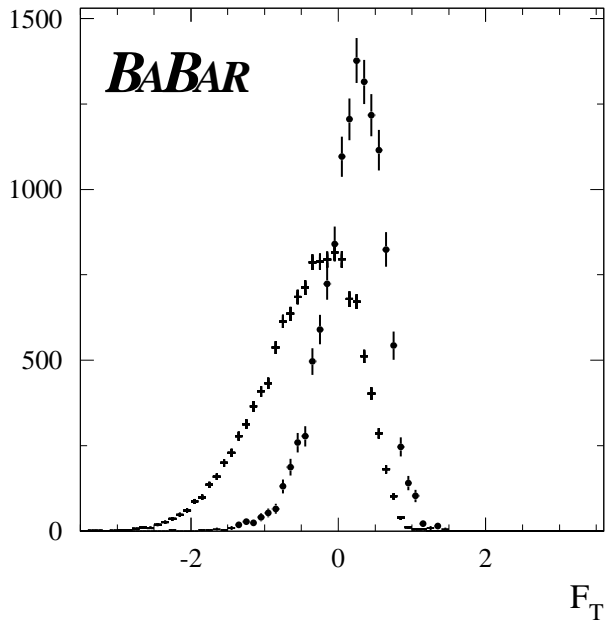}
\caption{\it Fisher discriminant output using the variables $L_0$ and $L_2$ 
(\cf\ Eq.~\ref{eq:monomialsApp}) for Monte Carlo $B^0\to\pi^+\pi^-$
signal (dots) and continuum background (crosses).
\label{fig:fisher} }
\end{figure}

\section{Charmless \B\ decays}

\subsection{Inclusive rates and direct CP violation in $B^+\to h^+h^-h^+$ ($h=\pi, K$)}
\label{sec:hhh}

Measurements of the $B^+\to h^+h^-h^+$ ($h=\pi, K$) decays can be used to determine the Unitarity
Triangle angle $\gamma$ and to help reducing uncertainties on the measurement
of the angle $\alpha$\cite{snyder}. Amongst the six final states considered here,
two of them do not occur in first or second order in the weak interaction coupling, and
are therefore highly suppressed in the Standard Model: $B^+ \to K^-\pi^+\pi^+$ and $B^+ \to K^+K^+\pi^-$. 

$B^+\to h^+h^-h^+$ decays have been studied using an integrated luminosity of 
$81.9 \ifb$. \B\ candidates consist of combinations of three charged tracks having at least $12$
hits in the DCH, a minimal transverse momentum of $100~\MeVc$ and originating from the beam spot.
The $\de$ and $\mes$ variables are computed assuming the three tracks to be pions. For modes
containing kaons, the resulting $\de$ distributions are shifted by roughly
$-45$ MeV per kaon. Charged pions and kaons are identified using $dE/dx$ information
from the SVT and the DCH, and for tracks with momentum above $700~\MeVc$, using the Cherenkov
angle and number of photons measured by the DIRC. Kaon selection efficiency is on
average $80\%$ and mis-identification of pions as kaons is below $5\%$ up to a momentum
of $4~\GeVc$. Pions are required to fail both the kaon and electron selection algorithms.
The latter is based on $dE/dx$, EMC shower shapes, and $E/p$ ratio. The mis-identification
of electrons as pions is $5\%$, and of kaons as pions is $20\%$.

Candidates with an intermediate neutral resonance mass compatible with any of the charm meson
$D^0$, $J/\psi$, $\psi(2S)$ and $\chi_{c,0}$ masses are vetoed.
Continuum background is suppressed cutting on $\cosBthr$ and a Fisher discriminant formed from the summed
scalar momenta of all charged and neutral particles from the rest of the event within the nine
nested cones coaxial with the thrust axis of the \B\ candidate\cite{CLEOFisher} (we will
refer to this Fisher discriminant as the CLEO Fisher in the following).
The cuts on these topological variables are optimized for each signal mode to achieve
maximum sensitivity for the branching fraction, and lead to a rejection of around $90\%$
of the continuum background. The residual background level is extrapolated from the sideband
region in the $\mes-\de$ plane into the signal region (defined as $|\mes-m_B|<8~\MeVcd$ and
$|\de-\langle\de\rangle|<60$ MeV where $\langle\de\rangle|$ is the mean value of
$\de$ measured in the control sample $B^-\to D^0\pi^-$ and $B^-\to D^0 K^-$). 

Several effects are taken care of when computing the branching fraction of each signal mode:
\begin{itemize}
\item The dependence of the efficiency of the signal selection described above 
on the position of the event in the Dalitz plot. 
\item The Dalitz plot dependent cross-feed between modes with $N$ kaons towards modes with $N-1$ kaons due to
mis-identification of kaons as pions.
\item The Dalitz plot independent cross-feed due to double mis-identification of kaons as pions, or
mis-identification of pions as kaons.
\item The Dalitz plot independent remaining charm and charmless \B\ background cross-feed coming
from $D^0$ and $\overline{D}^0$ for the $B^+\to\pi^+\pi^-\pi^+$ and $B^+\to K^+\pi^-\pi^+$ channels, 
and from $B^+\to \eta^\prime(\to \rho^0\gamma)K^+$ for the $B^+\to K^+\pi^-\pi^+$ channel. 
\end{itemize}

Results are summarized in Table~\ref{tab:results}. Signals are observed with around
$6\sigma$ of significance for the modes $B^0 \to \pi^+\pi^-\pi^+$, $B^0 \to K^+\pi^-\pi^+$
and $B^0 \to K^+K^-K^+$. For these modes, both the branching ratio and the direct
CP asymmetry are quoted. All \cp\ asymmetries are compatible with zero.
The significance of the $B^0 \to K^+K^-\pi^+$ signal is weak,
therefore only a $90\%$ CL upper limit is quoted. Finally, no events
are observed for the two Standard-Model highly suppressed decays $B^+ \to K^-\pi^+\pi^+$ 
and $B^+ \to K^+K^+\pi^-$. 

Systematic uncertainties on the branching fractions arise from background and cross-feed
estimations, and from the signal efficiencies (charged particle tracking, topological
variables cuts, particle identification, $\de$ and $\mes$ variables). Systematic
uncertainties on the CP asymmetries arise from tracking charge bias and particle identification. 

\subsection{Exclusive branching fractions of $B^+\to K^+\pi^-\pi^+$}
\label{sec:kpp}
The study of the $B^+\to K^+\pi^-\pi^+$ decay aims at various goals:
\begin{itemize}
\item search for direct CP violation,
\item constrain the angle $\gamma$ using interferences between $B^+\to\chi_{c0}K^+$
and other $B^+\to K^+\pi^-\pi^+$ decays\cite{gammaKpp,gammaKpp2},
\item determine the contributions from resonances involved.
\end{itemize}

In the analysis presented here, the branching ratios of both resonant and non-resonant 
$B^+\to K^+\pi^-\pi^+$ decays are measured in two steps:
\begin{itemize}
\item in a first step, the $B^+\to K^+\pi^-\pi^+$ Dalitz plot is split into eight regions which are
expected to be dominated by a particular resonance. The yield in each region is
measured using a maximum likelihood fit, with no assumption on the intermediate resonance.
\item in a second step, these yields are interpreted as branching fractions assuming a model
for the contributions to the Dalitz plot. The uncertainties on this model and the effects of overlap
and interferences between the various contributions are considered in the systematic studies. 
\end{itemize}

\subsubsection{Yield measurement in each Dalitz regions}

The Dalitz regions are described in Table~\ref{tab:splitdalitz}. Regions I, II and III (resp. IV, V and 
VI) are narrow bands in the invariant mass \mkp (resp. \mpp). 
The resonances contributing to the regions II and VI ``higher'' modes are unknown at this stage.
The ``high mass'' region VII could contain higher charmless and charmonium resonances
as well as a non-resonant contribution.

The areas where the \mkp narrow bands cross the $\pi\pi$ resonances are excluded to avoid 
interferences, and similarly where the $\pi\pi$ resonances cross the $\overline{D}^0$ band.
The other crossing regions are not excluded as the integrated interferences vanish (as long as the
\mkp\ cuts are symmetric). 

Finally, the charm mode region III is used as a control sample for systematic studies. 

\begin{table}[hbtp]
\centering
\caption{ \it Regions in the $B^+\to K^+\pi^-\pi^+$ Dalitz plot. The symbols ``!$\chi_{c0}$'' and
``!$\overline{D}^0$'' imply the exclusion of the $\chi_{c0}$ resonance ($3.355<$\mpp$<3.475$ \GeVcc), 
and the $\overline{D}^0$ ($1.8<$\mkp$<1.9$ \GeVcc).}
\vskip 0.1 in
{\footnotesize
\begin{tabular}{llcc}
\hline
            & Dominant     & \multicolumn{2}{c}{Selection criteria} \\
\rs{Regions} & contribution & \mkp (\GeVcc) &  \mpp (\GeVcc)  \\
\hline
I   & $K^{*0}(892)\pi^+$    & 0.816 $<$ \mkp $<$ 0.976 & \mpp $>$ 1.5 and !$\chi_{c0}$ \\
II  & higher $K^{*0}\pi^+$  & 0.976 $<$ \mkp $<$ 1.8 & \mpp $>$ 1.5 and !$\chi_{c0}$ \\
III & $\overline{D}^0\pi^+$ & 1.835 $<$ \mkp $<$ 1.895 &  ~!$\chi_{c0}$ \\
IV  & $\rho^0(770)K^+$      & !$\overline{D}^0$ & 0.6 $<$ \mpp $<$ 0.9 \\
V   & $f_0(980)K^+$         & !$\overline{D}^0$ & 0.9 $<$ \mpp $<$ 1.1 \\
VI  & higher $fK^+$         & !$\overline{D}^0$ & 1.1 $<$ \mpp $<$ 1.5 \\
VII & higher mass           & \mkp $>$ 1.9 & \mpp $>$ 1.5 and !$\chi_{c0}$ \\
VIII & $\chi_{c0}K^+$       & \mkp $>$ 1.9 & 3.37 $<$ \mpp $<$ 3.46 \\
\hline
\end{tabular}
}
\label{tab:splitdalitz}
\end{table}

Signal events are selected by forming three charged track combinations where two tracks
are identified as pions, and one as kaon using the methods described in
section~\ref{sec:hhh}. 

Continuum background is suppressed by requiring $|\cosBthr|<0.9$ and by using the CLEO Fisher
in the maximum likelihood fit. 

\B\ backgrounds arise from the following sources: combinatorial background from unrelated tracks;
specific three- and four-body $B\to D$ decays; charmless three- and four-body decays (mainly
$B^+\to\eta^\prime(\rho^0(770)\gamma)K^+$).
The backgrounds which significantly contribute to the signal yields are parameterized in 
the final fit, otherwise, they are subtracted from the signal yield. The modes
$B^+\to J/\psi K^{(*)+}$ and $B^+\to\psi(2S)K^{(*)+}$ are vetoed. 

A maximum likelihood fit is performed to extract the yield in each Dalitz region. 
Probability Density Functions (PDFs) are formed for the variables $\mes$, $\de$ and
Fisher ${\cal F}$. The likelihood in each Dalitz region is given by:
\begin{equation}
{\cal L} = \exp\left( -\sum_{i=1}^{M}n_i \right)\prod_{j=1}^{N}\left( \sum_{l=1}^{M}n_l{\cal P}_l(\vec{\alpha},
\vec{x_j})\right),
\end{equation}
where ${\cal P}_l$ are the PDFs of the variables $\vec{x}_j=\{\de,\mes,{\cal F}\}$, parameterized
by the parameters $\vec{\alpha}$ (determined before the final multivariate fit), 
for the event number $j$ and the hypothesis
$l=\{$signal, continuum background, \B\ background$\}$. 

Figure~\ref{fig:dalitz} shows the Dalitz plot for on-resonance data within the signal region
$5.2715 < \mes < 5.2865$ \GeVcc~after cutting on a likelihood ratio formed from
the $\de$ and ${\cal F}$ PDFs to enhance the signal over background ratio. 

\begin{figure}[t]
\vspace{9.5cm}
\includegraphics{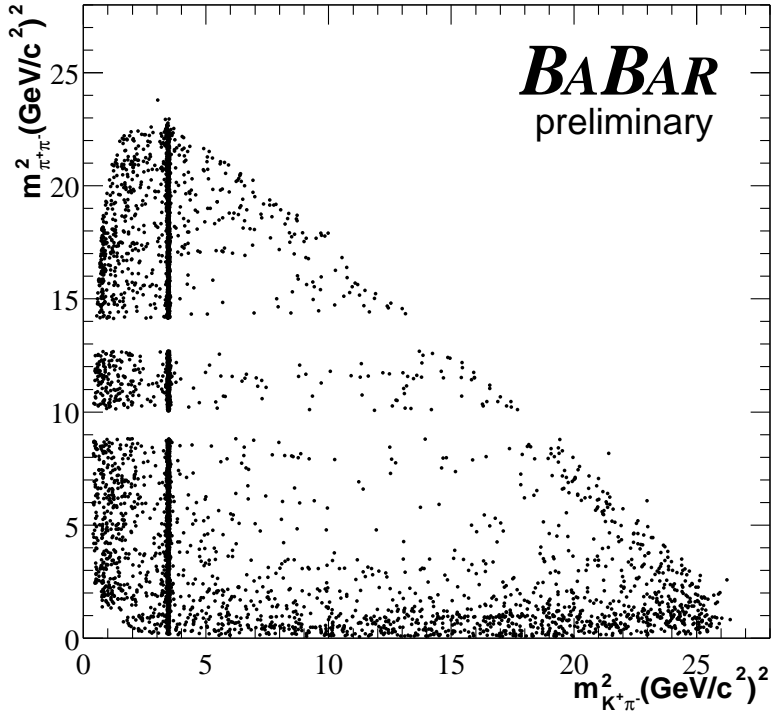}
\caption{\it Dalitz plot for on-resonance data within the signal region
$5.2715 < \mes < 5.2865$ \GeVcc~after cutting on a likelihood ratio formed from
the $\de$ and ${\cal F}$ PDFs to enhance the signal over background ratio}
\label{fig:dalitz}
\end{figure}

\subsubsection{Branching fraction measurement}

One distinguishes two categories of modes:
\begin{itemize}
\item modes which do not suffer from cross-feed with other modes, like
in regions III ($\overline{D}^0\pi^+$) and VIII ($\chi_{c0}K^+$). In this case, 
their branching ratios are simply obtained by:
\begin{equation}
{\cal B} = \frac{Y}{N_{B\bar{B}}\epsilon},
\end{equation}
where $Y$ is the signal yield measured in the previous stage, $N_{B\bar{B}}=(61.6\pm 6.8)\times 10^6$
is the number of $B\bar{B}$ pairs, and $\epsilon$ is the reconstruction efficiency calculated
using signal Monte Carlo events and corrected to account for data/Monte Carlo discrepancies
in tracking and particle identification.
\item for the other modes suffering from cross-talk, one uses:
\begin{equation}
{\cal B} = M^{-1}Y/N_{B\bar{B}},
\end{equation}
where ${\cal B}$ and $Y$ are now branching ratio and yield vectors, and $M$ is a matrix
representing the probability of an event of a particular mode to be found in a given
region. The branching fractions depend on the resonance model assumed in calculating
the matrix M, which is split into two component matrices, $P$ and $\epsilon$, such
that $M_{ij}=P_{ij}\epsilon_{ij}$. The $P$ matrix accounts for the event distribution within the 
Dalitz plot, and $\epsilon$ for the reconstruction efficiencies. One assumes one dominant resonance
per region, as indicated in the second column of table~\ref{tab:splitdalitz}. For regions II, VI and
VII, many contributions are possible: the model chosen here respectively includes $K_0^{*0}(1430)$, 
$f_2(1270)$ and a flat non-resonant $K^+\pi^-\pi^+$. The masses and widths are taken from
the PDG\cite{PDG}, and resonances are modelled by a Breit-Wigner, except for the $\rho(770)$
where the Blatt-Weisskopf parameterization is used\cite{BlattW}. This model suffers from large
uncertainties: the dominant resonance is unknown is some regions, there are uncertainties on
the masses and widths of the resonances and on the choice of lineshapes. Alternative
resonances and lineshapes are used for systematic studies. Possible interferences between 
resonances are also considered for systematics. 
\end{itemize}

The obtained branching fractions are given in table~\ref{tab:results}. The systematic errors
include components from the measurement itself (tracking efficiencies, particle identification,
Fisher PDF, number of $B\bar{B}$ pairs), from the model (resonances contributions, masses, widths
and lineshapes) and from the possible interferences (computed by allowing each contribution
to have a random phase). These two last sources of systematics are dominant over the first one,
except for the $B^+ \to \overline{D}^0\pi^+$ mode where only the first source is present.
Significant signals are observed in the
$B^+ \to K^{*0}(892)\pi^+$, $B^+ \to f_0(980)K^+$, $B^+ \to \chi_{c0}K^+$,
$B^+ \to \overline{D}^0\pi^+$ and $B^+ \to {\rm higher}~K^{*0}\pi^+$ channels

\subsection{Branching ratios of $B^+\to \rho^0\rho^+$ and $B^+\to \rho^0 K^{*+}$}
\label{sec:rr}
The first decay described here, $B^+\to \rho^0\rho^+$, enters the isospin analysis of the $B\to\rho\rho$ modes 
which aims at measuring the angle $\alpha$ of the unitarity triangle. The second decay, 
$B^+\to \rho^0 K^{*+}$, is expected to be dominated by $b\to s$ loops where physics beyond
the Standard Model could enter. 

These two decays are reconstructed in the following sub-decays: $K^{*+}\to K^+\pi^0$, 
$K^0(\to K^0_S(\to \pi^+\pi^-))\pi^+$, $\rho^+\to\pi^+\pi^0$ and $\rho^0\to\pi^+\pi^-$. 
Charged tracks are reconstructed using the same criteria than in section~\ref{sec:hhh},
except for the $K^0_S$ candidates to allow for a displaced vertex. The $K^0_S$ candidates 
must satisfy $|m(\pi^+\pi^-)-m(K^0)|<12$ \MeVcc, with the cosine of the angle
between their reconstructed flight and momentum directions greater than $0.995$,
and the measured proper decay time greater than five times its uncertainty. 
Photons with a minimum energy of $30$ MeV are paired to form $\pi^0$'s, with a typical
invariant mass resolution of $7$ \MeVcc. One therefore selects $\pi^0$ by applying a 
$\pm 15$ \MeVcc\ interval around the nominal $\pi^0$ mass. The $K^*$ and $\rho$
resonances are formed by pairing two particles which invariant masses are 
within the following intervals: $0.75 < m(K\pi) < 1.05$ \GeVcc\ for the $K^*$
and $0.52 < m(\pi\pi) < 1.00$ \GeVcc\ for the $\rho$. To suppress combinatorial
background, one cuts at $-0.5$ on the helicity angle, defined as the
angle between the direction of one of the two daughters ($K$ for $K^*$ and $\pi^+$
for $\rho$) and the parent \B\ direction in the resonance rest frame. 

\B\ mesons are kinematically isolated by requiring $\mes > 5.2$ \GeVcc\ and $|\de|<0.2$ GeV.
The continuum background is rejected by cutting on $|\cosBthr|<0.8$. A Fisher discriminant is also
constructed (and used in the final likelihood fit) 
based on the nine cones of the CLEO Fisher, and the additional variables $\cosBthr$
and the cosine of the polar angle between the \B\ momentum and the beam axis. 

Charmed \B\ background coming from $D\to K\pi, K\pi\pi$ is vetoed. The remaining small
\B\ background is accounted for in the fit. 

The final result is extracted using a maximum likelihood fit. The selection efficiencies for 
transverse and longitudinal angular polarization are averaged and are assigned a systematic error
defined by the RMS of a uniform efficiency between the extreme cases ($9\%$ for $\rho^0 K^{*+}_{K^0\pi^+}$,
$19\%$ for $\rho^0 K^{*+}_{K^+\pi^0}$, $18\%$ for $\rho^0 \rho^+$). The branching 
fractions are given in table~\ref{tab:results}~\footnote{The branching fractions
of $B^+\to \rho^0\rho^+$ and $B^+\to \rho^0 K^{*+}$ have been updated~\cite{LP03rhorho} 
since the preliminary results presented at this conference and read 
${\cal B}(B^+\to \rho^0\rho^+) = (22.5 ^{+5.7}_{-5.4}\pm 5.8)\times 10^{-6}$
and ${\cal B}(B^+\to \rho^0K^{*+}) = (10.6^{+3.0}_{-2.6}\pm2.4)\times 10^{-6}$.
The large difference between the preliminary and final results for $B^+\to \rho^0\rho^+$
is due to the high dependency of the selection efficiency on the polarization of the final
state. The preliminary analysis assumed a $50\%$-longitudinal / $50\%$-transverse 
polarization, whereas the final analysis measures a $100\%$-longitudinal polarization.}
Significant signals (above $4\sigma$) are observed in both channels. 

Projections plots of the invariant masses of $K\pi$ for the $B^+\to \rho^0 K^{*+}$ decay,
and $\pi^+\pi^-$ for the $B^+\to \rho^0\rho^+$ decay, are shown in Figure~\ref{fig:rr}.

\begin{figure}[t]
\vspace{8.0cm}
\includegraphics{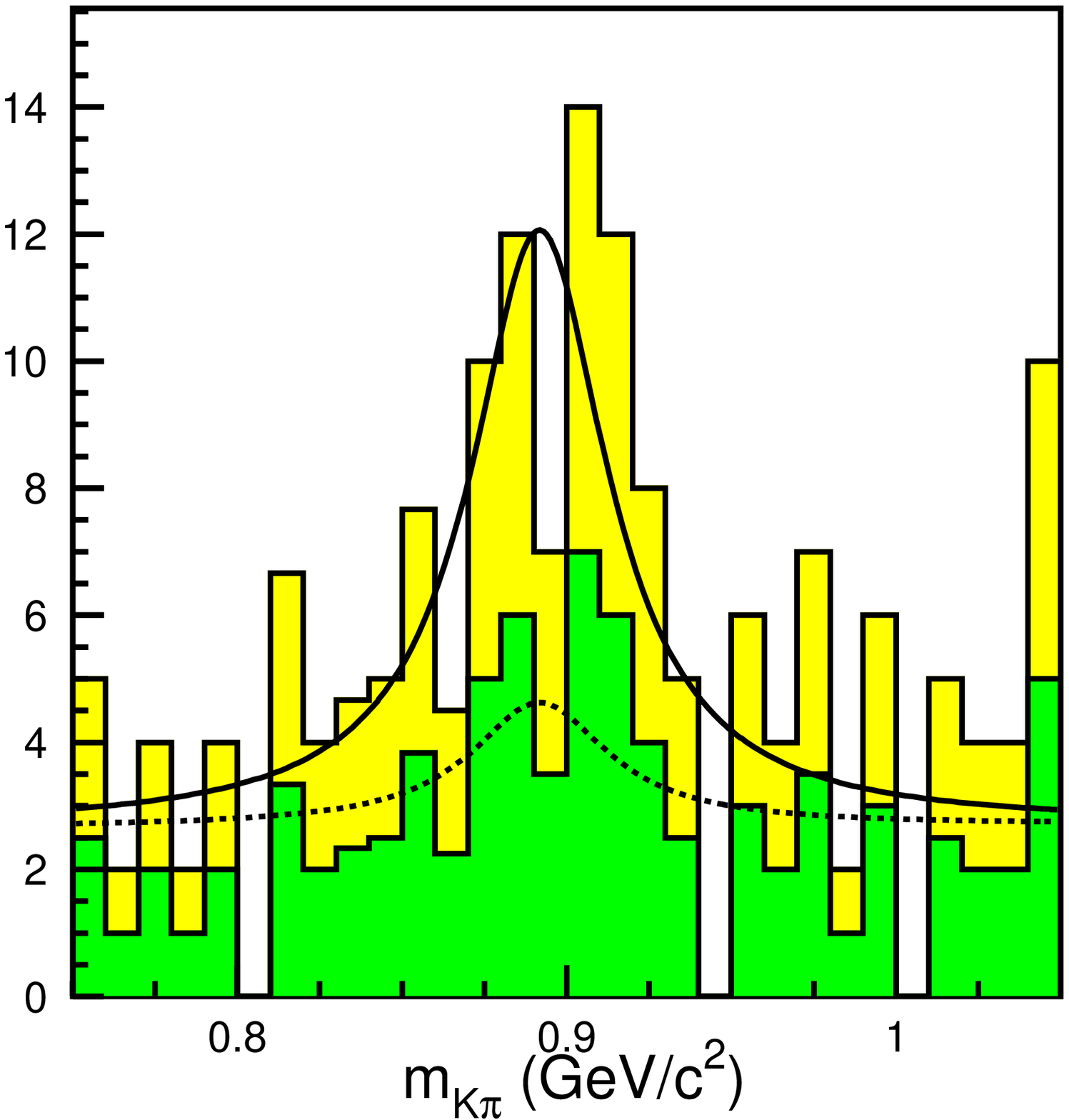}
\includegraphics{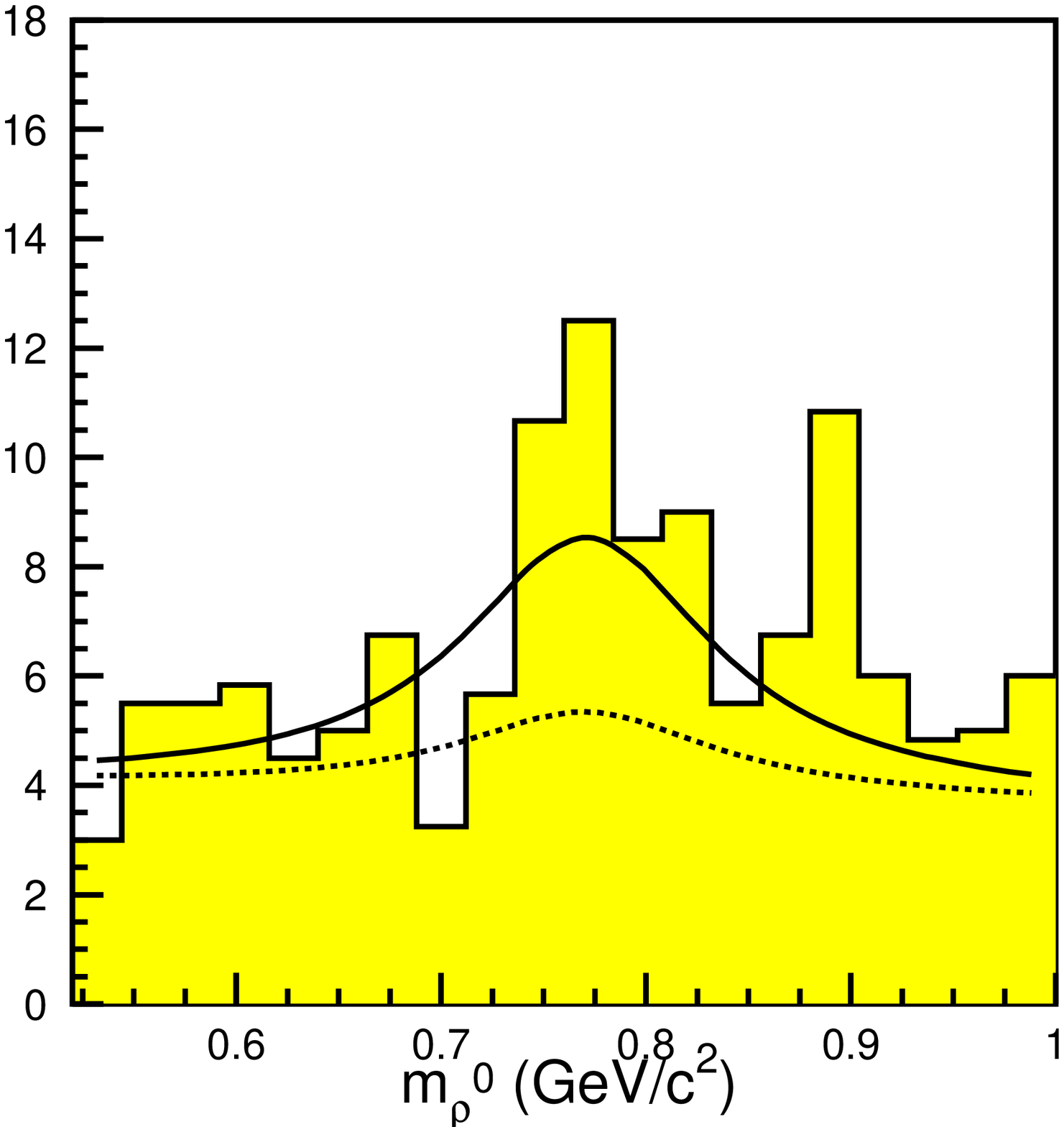}
\caption{\it  Projections onto the invariant masses of $K\pi$ (left plot)
for the $B^+\to \rho^0 K^{*+}$ decay, and of $\pi^+\pi^-$ (right plot)
for the $B^+\to \rho^0\rho^+$ decay. The histograms show the data (for the
$K^*$ projection, the shaded area shows the $K^+\pi^0$ final state only), and the solid
(resp. dashed) line shows the signal-plus-background (resp. background only) PDF projections. 
\label{fig:rr}}
\end{figure}

\section{Gluonic penguins}

\subsection{Branching fractions, longitudinal components, and direct \cp\ asymmetries in $B\to \phi K^*$}
\label{sec:phikst}

The decays $B\to \phi K^*$ are expected to proceed through pure $b \to s$ loops, where
new physics could enter. Therefore, the measurement of direct \cp\ violation, 
as well as the measurement of $\sin2\beta$ via the time-dependent analysis
of $B^0\to \phi K^{*0}$\cite{faccini}, can probe physics beyond the Standard Model.
The analysis of angular distributions in these vector-vector final states is also of
interest since information about the decay dynamics can be obtained\cite{vecvec}.

The analysis proceeds in a very similar way than what is described in section~\ref{sec:rr},
with the additional feature that longitudinal components and direct \cp\ asymmetries
are measured in the final likelihood fit. Due to limited statistics, the angular
analysis is simplified, and only the longitudinal components $f_L=\Gamma_L/\Gamma$
are obtained using two-helicity angle distributions:
\begin{equation}
\frac{1}{\Gamma}\frac{d^2\Gamma}{d\cos\theta_1d\cos\theta_2} = \frac{9}{4}
\left( \frac{1}{4}(1-f_L)\sin^2\theta_1\sin^2\theta_2 + f_L\cos^2\theta_1\cos^2\theta_2 \right),
\end{equation}
where $\theta_1$ and $\theta_2$ are respectively the helicity angles of the $K^*$ and the $\phi$. 

The results are summarized in table~\ref{tab:results}. Significant signals ($>10\sigma$)
are observed in both channels. Both direct \cp\ asymmetries are compatible with zero.

\begin{figure}[t]
\vspace{7.5cm}
\includegraphics{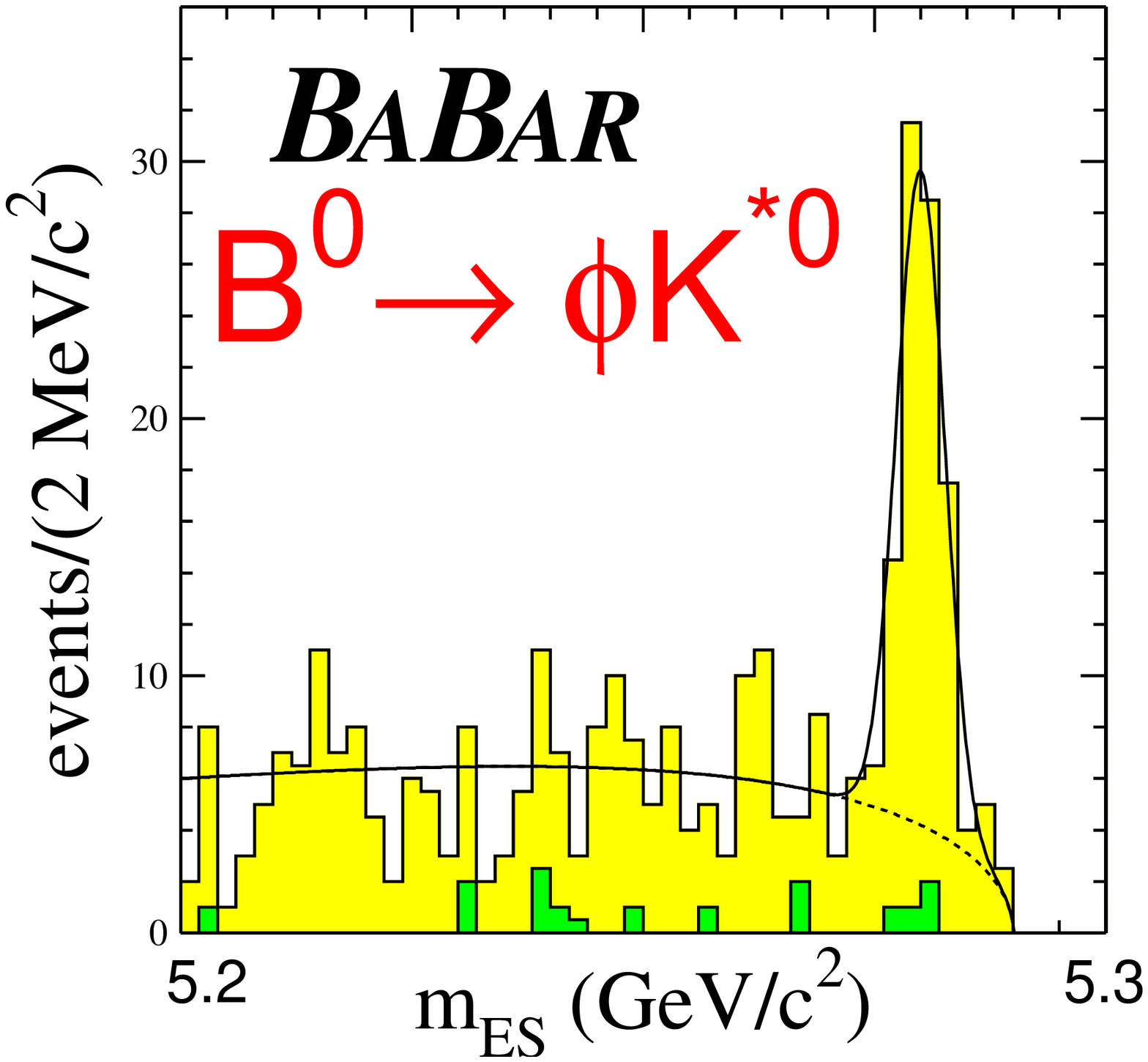}
\includegraphics{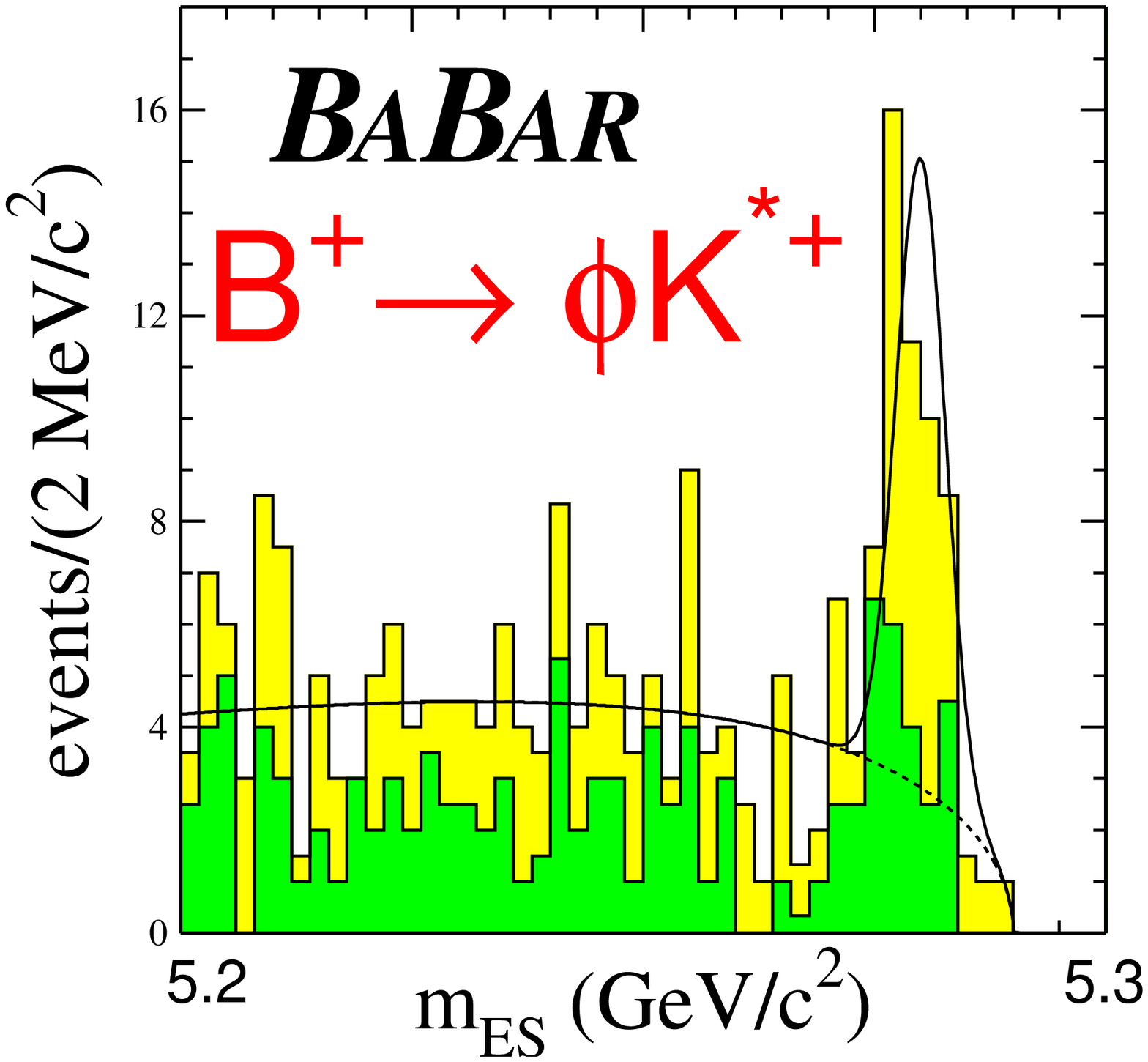}
\caption{\it  Projections onto $\mes$ for $B^0\to\phi K^{*0}$ (left plot) and
$B^+\to\phi K^{*+}$ (right plot). The histograms show the data (the shaded area shows 
the $K^+\pi^0$ final state only), and the solid (resp. dashed) line shows the 
signal-plus-background (resp. background only) PDF projections. 
\label{fig:phikst}}
\end{figure}

Projections onto $\mes$ of both signals are shown in figure~\ref{fig:phikst}. 

\section{Conclusions}

\begin{table}[t]
\centering
\caption{ \it Results for the measurements summarized in this proceeding. 
The first error quoted is statistical and the second error corresponds to
systematics. Upper limits are quoted at $90\%$ CL. }
\vskip 0.1 in
{\normalsize
\begin{tabular}{lccc}
\hline
\hline
& & & \\
\rs{Mode} & \rs{BF ($10^{-6}$)} & \rs{$A_{\rm CP}$} & \rs{$\cal L$ ($\ifb$)}\\
\hline
\hline
& & & \\
\multicolumn{4}{c}{\rs{Charmless \B\ decays}} \\
\hline
& & & \\
\multicolumn{4}{c}{\rs{Inclusive $B^+\to h^+h^-h^+$ ($h=\pi, K$) (sec.~\ref{sec:hhh})}} \\
$B^+ \to \pi^+\pi^-\pi^+$ & $10.9 \pm 3.3 \pm 1.6$ & $-0.39 \pm 0.33 \pm 0.12$ & 81.9 \\
$B^+ \to K^+\pi^-\pi^+$   & $59.1 \pm 3.8 \pm 3.2$ & $\phantom{1}0.01 \pm 0.07 \pm 0.03$ & 81.9 \\
$B^+ \to K^+K^-K^+$       & $29.6 \pm 2.1 \pm 1.6$ & $\phantom{1}0.02 \pm 0.07 \pm 0.03$ & 81.9 \\
$B^+ \to K^+K^-\pi^+$     & $< 6.3$ & - & 81.9 \\
$B^+ \to K^-\pi^+\pi^+$   & $< 1.8$ & - & 81.9 \\
$B^+ \to K^+K^+\pi^-$     & $< 1.3$ & - & 81.9 \\
\hline
& & & \\
\multicolumn{4}{c}{\rs{Exclusive $B^+\to K^+\pi^-\pi^+$ (sec.~\ref{sec:kpp})}} \\
$B^+ \to K^{*0}(892)\pi^+$ & $10.3 \pm 1.2~^{+1.0}_{-2.7}$ & - & 56.4\\
$B^+ \to f_0(980)K^+$      & $9.2 \pm 1.2~^{+2.1}_{-2.6}$  & - & 56.4\\
$B^+ \to \chi_{c0}K^+$     & $1.46 \pm 0.35 \pm 0.12$     & - & 56.4\\
$B^+ \to \overline{D}^0\pi^+$ & $184.6 \pm 3.2 \pm 9.7$   & - & 56.4\\
$B^+ \to {\rm higher}~K^{*0}\pi^+$ & $25.1 \pm 2.0~^{+11.0}_{-5.7}$ & - & 56.4\\
$B^+ \to \rho^0(770)K^+$   & $< 6.2$ & - & 56.4\\
$B^+ \to K^+\pi^-\pi^+$    & $< 17.0$ & - & 56.4\\
$B^+ \to {\rm higher~}f K^+$ & $<12.0$ & - & 56.4\\
\hline
& & & \\
\multicolumn{4}{c}{\rs{$B^+\to \rho^0\rho^+$ and $B^+\to \rho^0 K^{*+}$ (sec.~\ref{sec:rr})}} \\
$B^+\to \rho^0\rho^+$ & $9.9 ^{+2.6}_{-2.5} \pm 1.1$ & - & 81.9\\
$B^+\to \rho^0 K^{*+}$ & $7.7^{+2.1}_{-2.0}\pm 1.4$ & - & 81.9 \\
\hline
& & & \\
\multicolumn{4}{c}{\rs{Gluonic penguins}} \\
\hline
& & & \\
\multicolumn{4}{c}{\rs{$B\to \phi K^*$ (sec.~\ref{sec:phikst})}} \\
$B^0\to\phi K^{*0}$ & $11.1 ^{+1.3}_{-1.2}\pm 1.1$ & $+0.04 \pm 0.12 \pm 0.02$ & 81.9 \\
                    & \multicolumn{2}{c}{$\Gamma_L/\Gamma$ = $0.65 \pm 0.07 \pm 0.04$} & \\
$B^+\to\phi K^{*+}$ & $12.1 ^{+2.1}_{-1.9} \pm 1.5$ & $+0.16 \pm 0.17 \pm 0.04$ & 81.9 \\
                    & \multicolumn{2}{c}{$\Gamma_L/\Gamma$ = $0.46 \pm 0.12 \pm 0.05$} & \\
\hline
\hline
\end{tabular}}
\label{tab:results}
\end{table}

All results presented in these proceedings are summarized in table~\ref{tab:results}. 
Many new channels are observed with a large statistical significance. Measurements
of direct \cp\ violation do not indicate any significant deviation from zero. A first angular
analysis is performed on the vector vector final states $B\to \phi K^*$:
one measures a large longitudinal component, which should simplify the time
dependent analysis allowing to measure $\sin2\beta$ in $B^0\to \phi K^{*0}$. 
These results, and many others, should become extremely precise as the integrated
luminosity will reach $500\ifb$ in the year 2005, and more than $1~ab^{-1}$ at the
end of the decade.


\begin{thebibliography}{99}

\bibitem{babardet}   B.~Aubert \ea, \babar\ Collaboration, \Journal\NIMA{479}{1}{2002}
\bibitem{faccini}    R.~Faccini, these proceedings. 
\bibitem{Fisher}     R.A.~Fisher, Annals of Eugenics, {\bf 179} (1936)
\bibitem{snyder}     A.~Snyder and H.~Quinn, \Journal\PRD{48}{2139}{1993}
\bibitem{CLEOFisher} D.M.~Asner \ea, CLEO Collaboration, \Journal\PRD{53}{1039}{1996}
\bibitem{gammaKpp}   N.G.~Deshpande, G.~Eilam, X.G.~He, J.~Trampetic, \Journal\PRD{52}{5354}{1995}
\bibitem{gammaKpp2}  S.~Fajfer, R.J.~Oakes, T.N.Pham, \Journal\PLB{539}{67}{2002}
\bibitem{PDG}	     Particle Data Group, \Journal\PRD{66}{01001}{2002}
\bibitem{BlattW}     J.M.~Blatt and V.F.~Weisskopf, {\it Theoretical Nuclear Physics}
		     (Wiley, New York, 1952) 361
\bibitem{LP03rhorho} B.~Aubert \ea, \babar\ Collaboration, {\it hep-ex}/0307026
\bibitem{vecvec}     G.~Kramer, W.F~Palmer, \Journal\PRD{45}{193}{1992};
		     C.H.~Chen, Y.Y.~Keum, H.N.~Li, \Journal\PRD{66}{054013}{2002}


\end{thebibliography}
\end{document}